\documentclass[aps,prc,twocolumn,floatfix,showpacs,a4paper,
nofootinbib,amsmath,amssymb]{revtex4-1}
\usepackage{graphicx}
\usepackage{color}
\usepackage{dcolumn}
\usepackage{bm}
\newcommand{\be}{\begin{equation}}
\newcommand{\ee}{\end{equation}}
\newcommand{\ba}{\begin{eqnarray}}
\newcommand{\ea}{\end{eqnarray}}
\newcommand{\bd}{\begin{displaymath}}
\newcommand{\ed}{\end{displaymath}}
\newcommand{\bea}{\begin{eqnarray}}
\newcommand{\eea}{\end{eqnarray}}
\newcommand{\di}{{\rm d}}
\renewcommand{\vec}[1]{\mbox{\boldmath$#1$}}

\begin{document}

\title{Global $\Lambda$ Polarization in high energy collisions}
\author{Yilong Xie$^1$, Dujuan Wang$^2$, and L\'aszl\'o P. Csernai$^1$}
\affiliation{$^1$Institute of Physics and Technology, University of Bergen,
Allegaten 55, 5007 Bergen, Norway\\
$^2$ School of Science, Wuhan University of Technology, 
430070, Wuhan, China
}

\begin{abstract}
With a Yang-Mills flux-tube initial state and 
a high resolution (3+1)D Particle-in-Cell 
Relativistic (PICR) hydrodynamics simulation, 
we calculate the $\Lambda$ polarization for different energies.
The origination of polarization in high energy collisions is
discussed, and we find linear impact parameter 
dependence of the global $\Lambda$ polarization.
Furthermore, the global $\Lambda$ polarization in our model 
decreases very fast in the low energy domain, and the decline curve 
fits well the recent results of Beam Energy Scan (BES) 
program launched by the STAR collaboration at the Relativistic
Heavy Ion Collider (RHIC). 
The time evolution of polarization is also discussed.  
\end{abstract}

\date{\today\ - v6}


\maketitle

\section{Introduction}

The nontrivial polarization effect in high energy 
collisions, since it was firstly observed in Fermilab 
with both polarized and unpolarized incident beam 
\cite{BHM76,Lea75}, had been raising people's  interest. 
The $\Lambda$ hyperon is well suited to measure the
polarization because through the decay
$\Lambda^0 \rightarrow p + \pi^-$ with proton carrying 
the spin information,
the $\Lambda$ becomes its own spin-analyzer. Afterwards, 
more experimental research had been 
launched continuously, including nucleon collisions and 
heavy ion collisions 
\cite{Anea84,AMS87,EJR94,AM93,IA06,IS06,BIA07}.
Theoretical studies have also been underway synchronously with the experiments 
\cite{LW05a,HHW11,BGT07,Bea15,ADP86,BCDG13,BH05,Bec13,My1,My2}.

These experiments had observed that, 
1) the $\Lambda$ polarization is perpendicular to the reaction plane, 
2) and increases with the $\Lambda$'s transverse momentum ($p_T$) 
and its Feynman-x, taken to be
$x_F=p_L/\sqrt{s}\ $ \cite{AMS87,EJR94,IA06}. 
However, no significant evidence was found to indicate the energy dependence
of the hyperon polarization, which we will discuss in this paper.

The $\Lambda$ polarization in experiments was measured through the 
angular distribution of emitted protons
in $\Lambda$'s rest frame:
\bea
\label{exp-pi}
\frac{dN}{dcos\theta} = (1+\alpha P cos \theta)/4\pi ,
\eea
\noindent where $\theta$ is the angle between the 
proton momenta $\vec{p}_p$ and the $\Lambda$'s spin $\vec{S}_{\Lambda}$, 
$P$ is the polarization amplitude, and the decay parameter 
$\alpha$ is taken to be $0.647 \pm 0.013$\ \cite{IS06,BHM76}.
To perform the measurement and calculation, it is crucial to 
determine the Reaction Plane (RP) and Center
of Mass (CM) of the participant system. Recently it was 
pointed out that in collider experiments the CM frame 
determination might not be accurate enough due to the nuclear 
fragmentation effects while the early fixed target experiments 
can get rid of this issue\ \cite{Eyyubova}.

From the experiments, theorist suggested that the hyperon
polarization originates 
from the initial substantial 
angular momentum, $\vec{L}$, in non central collisions, 
since the global polarization aligns with the orbital 
angular momentum. The initial angular momentum is 
dependent on impact parameter, or centrality percentage, taking a 
shape of quadratic function that  peaks around 9\% 
centrality percentage, as shown in Refs. \cite{BPR08,GCD08}. 
In the RHIC's Au+Au collisions at 62.4 GeV and 200 GeV, 
no centrality dependence of the global hyperon polarization 
was analyzed \cite{KHPB15}, due to the insignificant
polarization. Recently, stronger polarization signal was 
observed in RHIC's Beam Energy Scan (BES) program in the energy 
region below 100 GeV \cite{Lisa2016}. Therefore, 
 in this paper we will try to explore this issue again.

During the past decades, two different perspectives 
were developed for the transition mechanism from 
initial angular momentum to the final state hyperon 
polarization, i.e. the hydrodynamical perspective 
and partonic kinetic perspective. From the partonic 
micro-perspective, the initial angular momentum is 
transferred to the partons through the interaction 
of spin-orbit coupling in viscous QGP \cite{HHW11}, 
and then the global polarized quarks are recombined 
into hadrons, in which the Thomas precession of the quark 
spin was applied \cite{TDHM14}.

In the hydro- and thermo- dynamical description, the 
initial angular momentum is manifested in a longitudinal 
velocity shear, which, with small shear viscosity, results 
into a rotating system with substantial 
vorticity and even Kelvin-Helmholtz instability \cite{CDA12}. 
Assuming local equilibrium at freeze out and equipartition of 
the spin degree of freedom, Ref. \cite{BCDG13} put forward 
a polarization 3-vector for spin 1/2 particles and antiparticles 
based on the generalization of Cooper-Frye formula for particles 
with spin.

It was recently pointed out that the detailed balance 
of Cooper-Frye formula on Freeze-Out(FO) hypersurface requires a 
non-vanishing polarization in fluid before FO 
\cite{MTG17}. However, the absence of pre-FO polarization should not 
significantly effect the polarization calculation based on Ref. 
\cite{BCDG13}. 
One can calculate that, the spin of each baryon is 
$ L = \hbar/2  \approx 98.5 \: MeV\cdotp fm/c$.
As the polarization is between  1 - 10 \% at different
beam energies in the RHIC BES program, this gives
$L \approx 1 - 10 \: MeV \cdotp fm/c$ for the angular momentum
carried by one baryon.
On the other hand the total angular momentum is around \cite{CWC14}:
$L = 1.05 \times 10^4  \hbar = 205.8 \times 10^4 \: MeV\cdotp fm/c$.
This is distributed among a few hundred baryons
in semi-peripheral reactions at not too high energies, i.e. 
very few antibaryons, which gives an angular momentum
per baryon:
$L \approx  10^4 \: MeV\cdotp fm/c$.
This is 3 - 4 orders of magnitude bigger than the spin
angular momentum carried by one baryon in the vortical flow.
Therefore, even if 1 - 10\% of spins are already polarized before FO, carrying only one per mil of the total angular momentum, they will neither effectively impact the fluid dynamical evolution, nor significantly change the detailed balance during FO process, thus keeping the validity of 
the polarization 3-vector in Ref. \cite{BCDG13}.

Refs. \cite{Bec13,My1} applied this polarization 
3-vector to relativistic heavy ion collisions, to explore the 
momentum space distribution of $\Lambda$ polarization. However, 
the previously neglected second term
of the polarization formula,
which reflects the effect of system expansion, turned out to be 
not negligible.
 In this paper, we will compute the complete $\Lambda$ 
 polarization, including both the first and second term, 
 for the Au + Au collisions in the same energy domain as the 
 RHIC BES program.

\section{$\Lambda$ Polarization in Hydrodynamic model}

The initial state we used here could naturally generate 
a longitudinal velocity 
shear \cite{M1,M2}, which
leads to the hyperon polarization after the 
hydrodynamical evolution, 
simulated by a high resolution Computational
Fluid Dynamic (CFD) calculation using the 
Relativistic Particle-in-Cell (PICR) method. 
This initial state assumed a
Yang-Mills field string tension between Lorentz 
contracted streaks after impact, 
and  conserved the
angular momentum both locally and globally. 
Both in the initial state and subsequent CFD simulation, the
frequently used `Bag Model' EoS was applied: 
$P=c_0^2 e^2 - \frac43 B$, 
with constant $c_0^2 = \frac13$ and a fixed 
Bag  constant $B$ \  \cite{M1,M2,LPS94}.
The energy density takes the form:
$e=\alpha T^4+ \beta T^2 + \gamma + B $, where 
$\alpha$, $\beta$, $\gamma$ 
are constants arising from the degeneracy 
factors for (anti-)quarks and gluons. 
At Freeze-Out (FO) stage, the major part of  
FO hypersurface is assumed to be 
timelike, which entails small changes between 
the pre-FO and post-FO state, 
and thus the ideal gas phase space distribution 
can be applied \cite{ChengEtAl10,My1}.

The spatial part of polarization 3-vector for 
(anti-) hyperon with mass $m$, 
reads as \cite{Bec13,My1,My2}:
\bea
\label{Pipv}
 && \vec{\Pi}(p) = \frac{\hbar \varepsilon}{8m}
  \frac{\int \di \Sigma_\lambda p^\lambda \, n_F\ 
  (\nabla\times\vec{\beta})}{\int \di \Sigma_\lambda p^\lambda \,n_F} 
  \nonumber \\
 && + \frac{\hbar {\bf p}}{8m} \times 
  \frac{\int \di \Sigma_\lambda p^\lambda \,
 n_F\ (\partial_t \vec{\beta} + \nabla\beta^0)}
 {\int \di \Sigma_\lambda p^\lambda \,n_F} \ ,
\eea
where $\beta^{\mu}(x)=(\beta^0,\vec{\beta})=[1/T(x)]u^{\mu}(x)$
is the inverse temperature four-vector field, and 
$n_F(x,p)$ is the Fermi-J\"uttner distribution of the $\Lambda$, 
that is $1/(e^{\beta(x)\cdot{p}-\xi(x)}+1)$, being 
$\xi(x)=\mu(x)/T(x)$ with $\mu$ being the
$\Lambda$'s chemical potential and $p$ its four-momentum.  
$\di \Sigma_\lambda$ is the freeze out hypersurface element, for
$t=$const. freeze-out, 
$\di \Sigma_\lambda p^\lambda\ \to \di V \varepsilon$,
where $\varepsilon = p^0$ being the $\Lambda$'s energy.

Here the {\it first term} reflects the classical vorticity 
effect ($\nabla \times \beta$), and the 
{\it second term} arises from the expansion effect 
($\partial_t \vec{\beta}$) and relativistic modification ($\nabla\beta^0$). 
Noticing that the convention of $\Pi(p)$ is normalized to
50\%, i.e. Eq. (\ref{exp-pi}),  
the value should be multiplied by 2 to keep in 
line with the polarization anisotropy in experimental studies, 
where the upper limit is 100\%.
This is unlike the previous studies \cite{Bea15,Bec13,My1,My2}.
Besides, the Eq. (\ref{Pipv}) is calculated in the Center-of-Mass (CM) 
frame, and one can Lorentz boost it into $\Lambda$'s rest frame 
by the following formula:
\be
\vec{\Pi}_0(\vec{p})=
\vec{\Pi}( p )-\frac
{\vec{p}}
{p^0 (p^0 + m)}
\vec{\Pi}( p ) \cdot \vec{p} \ .
\label{Pi0}
\ee

\begin{figure}[ht] 
\begin{center}
      \includegraphics[width=7.4cm]{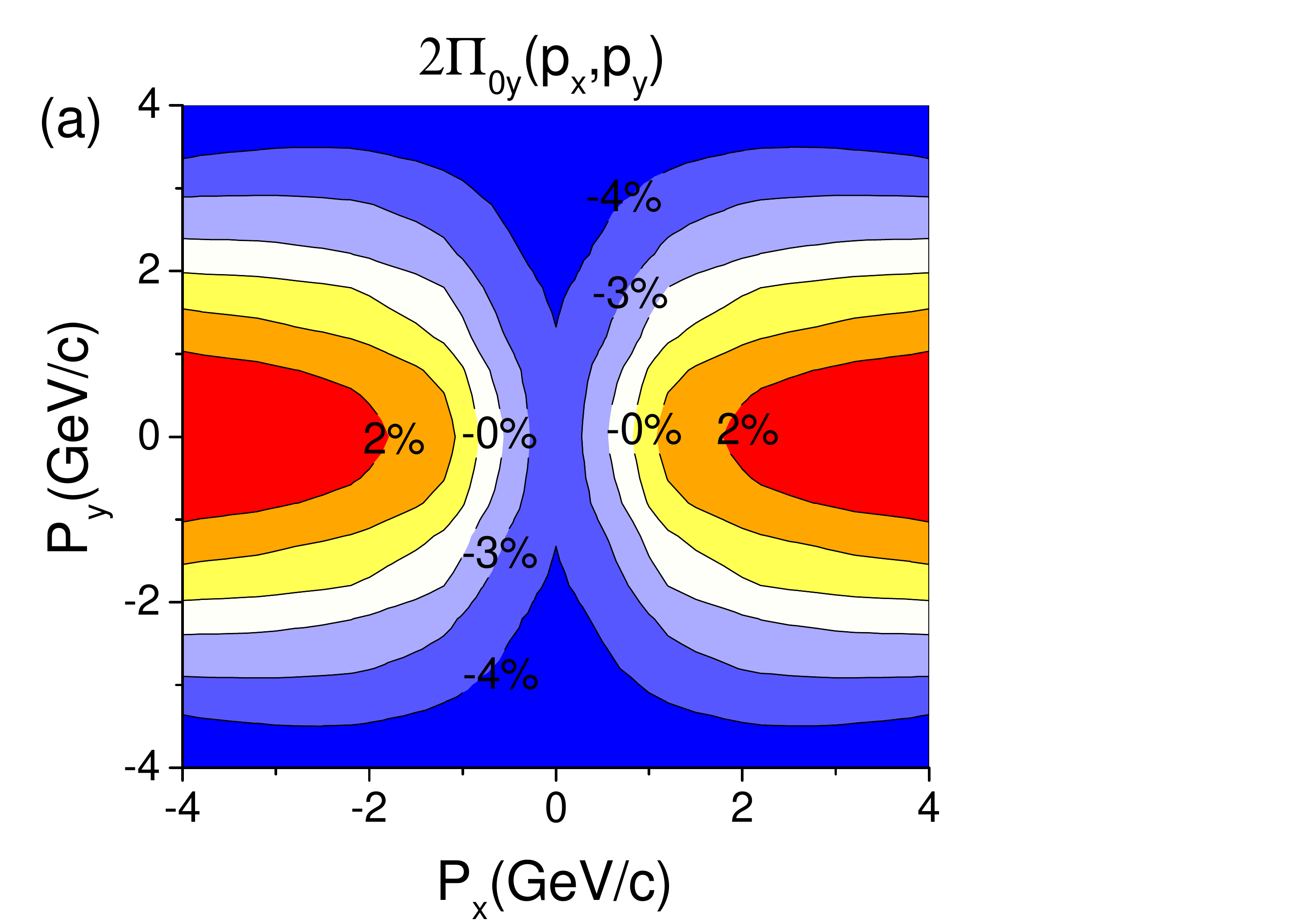}
      \includegraphics[width=7.4cm]{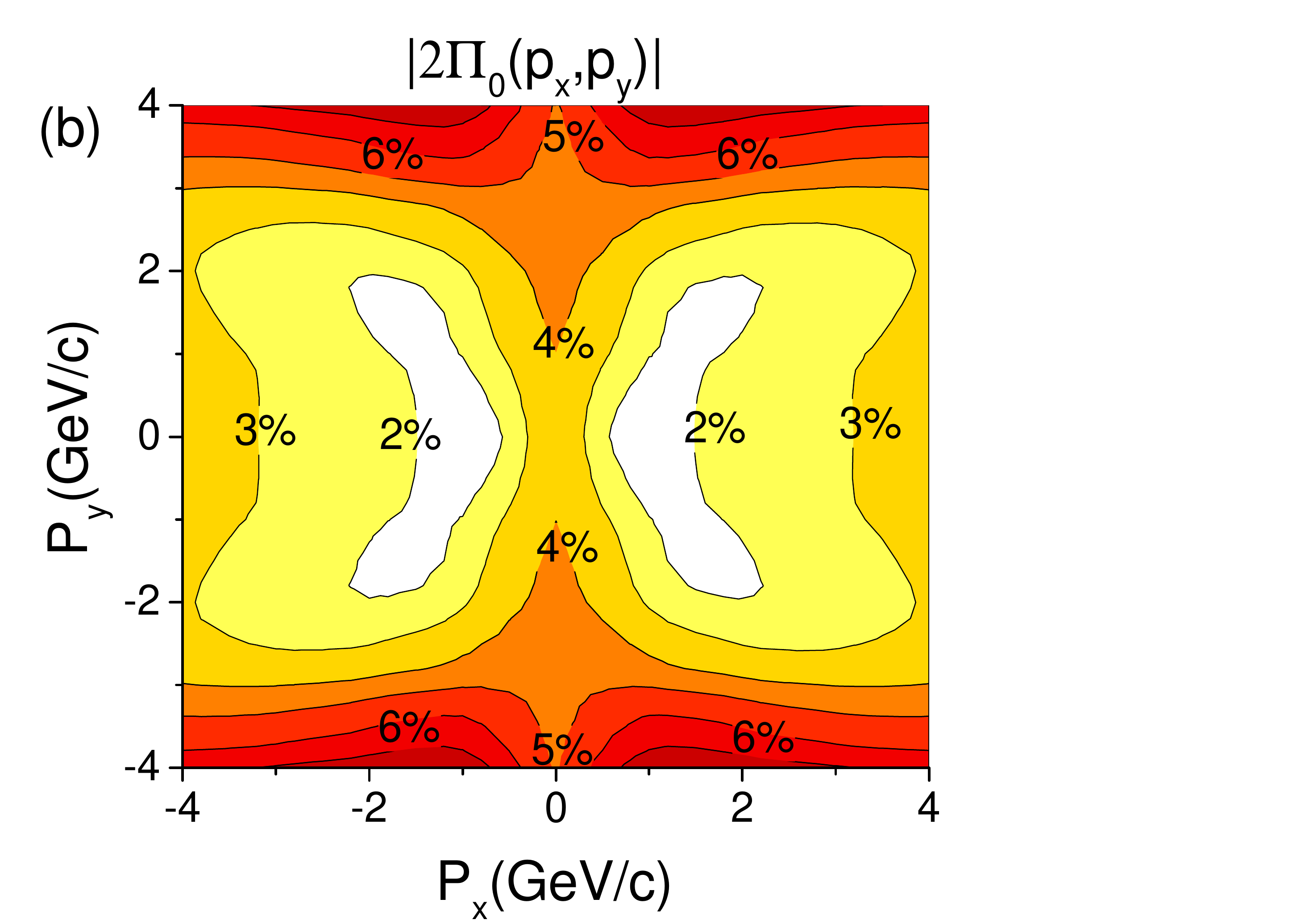}
\end{center}
\caption{
(Color online)
The $y$ component (top) and the modulus (bottom) of the $\Lambda$ 
polarization for momentum vectors in the transverse, 
$[p_x,p_y]$, plane at $p_z=0$, 
for the Au+Au reaction at $\sqrt{s_{NN}}=11.5$ GeV.
The figure is in the frame of the $\Lambda$. The impact
parameter $b=0.7 b_m = 0.7\times 2R$, where $R$ is the radius of 
Au and $b_m=2R$ is the maximum value of $b$.
The freeze out time is $6.25 =(2.5 + 4.75)$ fm/c, including
2.5 fm/c for initial state and 4.75 fm/c for hydro-evolution.
}
\label{F1}
\end{figure}

The three components of the polarization 3-vectors, 
$2\vec{\Pi} (p_x, p_y)$ (or $2\vec{\Pi_0} (p_x, p_y)$) 
have different significance.
As we  pointed out in our earlier paper \cite{My2},
the $x$ and $y$ components of polarization, $2\Pi_x$ and 
$2\Pi_x$, in transverse momentum space [$p_x, p_y$] 
are rather trivial and form a symmetric dipole structure, 
which results in vanishing global polarization along 
the $x$ and $y$ direction in the participant CM frame. Meanwhile, 
as expected, the  $-y$ directed polarization, aligned with the initial 
angular momentum, dominates the modulus of polarization 3-vector, 
$2\vec{|\Pi_0 (p_x, p_y)|}$.
Fig. \ref{F1} shows the dominant $y$ component and the modulus of
$\Lambda$ polarization, in Au-Au collisions at 11.5 GeV. 
One can see that the top and down figures
have similar structure and magnitude, which indicates a trivial influence 
of the $x$ and $y$ components on the global polarization.

Since the $- y$ directed global $\Lambda$ polarization 
in experimental results is
averaged polarization over the $\Lambda$'s momentum space, 
we evaluated the
average of the $y$ component of the polarization 
$\langle\Pi_{0y}\rangle_{p}$.
We integrated the  $y$ component of the obtained
polarization, $\Pi_{0y}$, over the momentum space as follows:
\ba
\langle\Pi_{0y}\rangle_{p}&=&
\frac{\int dp\, dx \,\Pi_{0y}(p,x) \, n_F(x,p)}{\int dp\, dx\, n_F(x,p)} 
\nonumber \\
&=&\frac{\int \,dp \,\Pi_{0y}(p)\, n_F(p) }{ \int \,dp \,n_F(p)}
\ea
to calculate the global polarization. The word `global' means averaging over 
phase space [$\vec{x}$, $\vec{p}$]. 
Besides, we replace the $\langle\Pi_{0y}\rangle_{p}$ 
with $-\langle\Pi_{0y}\rangle_{p}$, since in experiments the 
angular momentum's direction, i.e. negative $y$ direction is 
the conventional direction for global polarization.

\section{Results and Discussion}

\subsection{Angular momentum, Impact Parameter and Centrality}

According to 
the alignment of polarization and the system's angular momentum, 
theorists suggested to
attribute the polarization to the initial orbital angular 
momentum arising in non-central collisions. 
Refs. \cite{BPR08,GCD08} have analytically deduced and 
schematically displayed the initial angular momentum  in the 
reaction region as a  function of impact parameter $b$, taking 
the form of quadratic 
function, which roughly peaks at $b = 0.25 b_m$ or $0.3 b_m$. 
If the angular momentum is 
translated into polarization without any other significant 
perturbative mechanism, one 
should also observe the polarization's dependence on impact parameter.
In other words, the initial angular momentum of the participant system 
is initiated by the inequality of local nuclear density in the 
transverse plane, and this inequality is dependent on the impact 
parameter. Thus the initial impact parameter dependence of the 
late-state polarization should in principle be observed. 

\begin{figure}[ht] 
\begin{center}
      \includegraphics[width=7.4cm]{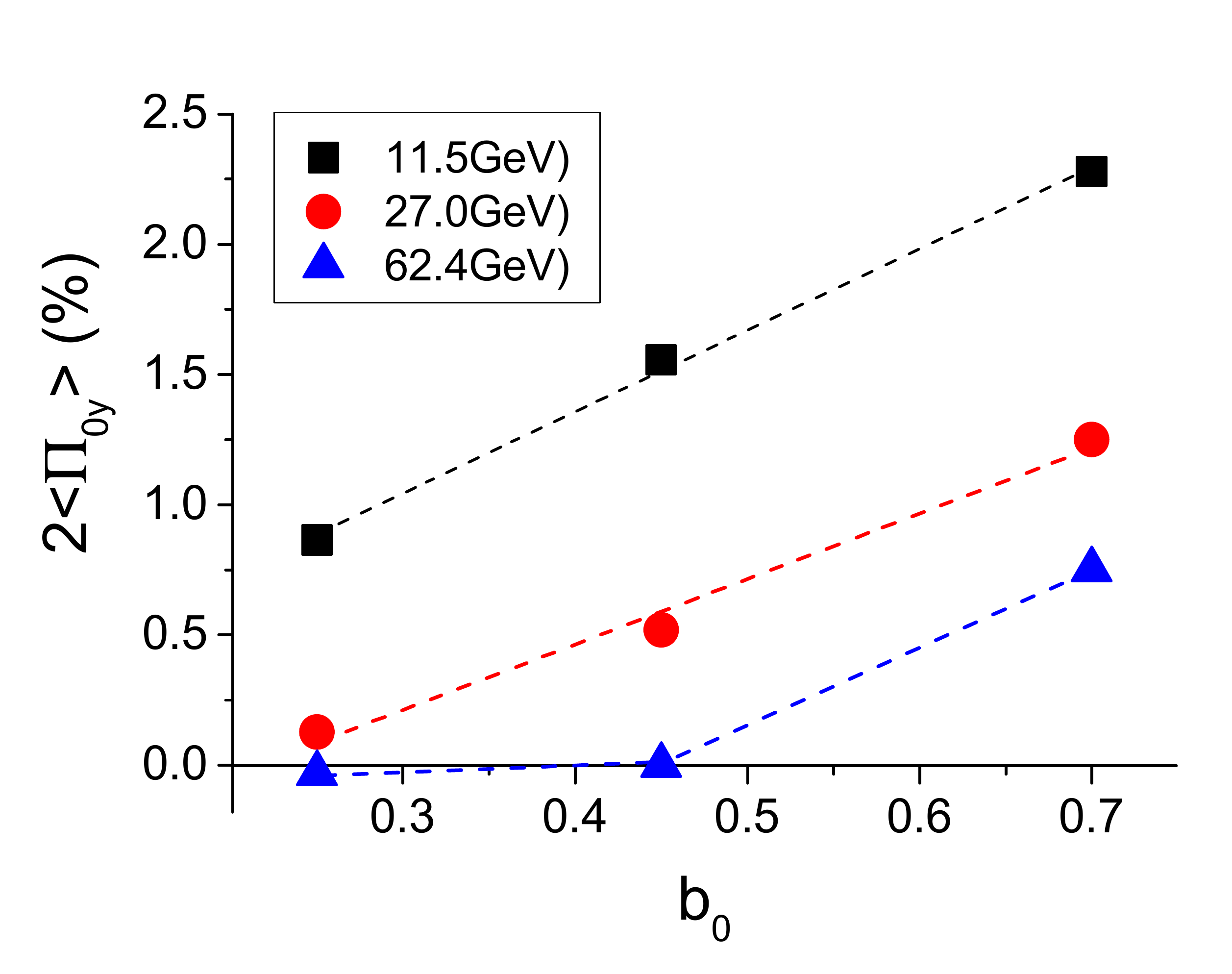}
\end{center}
\caption{
(Color online) The linear dependence of global polarization, 
$2\langle\Pi_{0y}\rangle_{p}$, 
as a function of impact parameter ratio $b_0$ at 
11.5 GeV, 27.0 Gev and 62.4 GeV. 
{\color{red}
}
}
\label{F2}
\end{figure}

Fig. \ref{F2} shows the global polarization of 
Au+Au collisions as a function of ratio of impact parameter 
$b$ to Au's nuclear radius $R$, 
i.e. $b_0 = b/2R$. 
One could see that 
the polarization at different energies indeed approximately 
takes a linear increase 
with the increase of impact parameter, except for 62.4GeV due 
to the vanishing polarization signals at relatively central 
collisions. This linear dependence clearly indicates that 
the polarization in our model arises from the initial angular 
momentum. However, the 
polarization's linear dependence on $b$ is somewhat different 
from the angular 
momentum's quadratic dependence on $b$. This is because the 
angular momentum $L$ is 
an extensive quantity dependent on the system's mass, while 
the polarization $\Pi$ is 
an intensive quantity. 


%
%


An earlier $\Lambda$ global polarization measurement 
by STAR in Au+Au collisions at
62.4 GeV and 200 GeV, had observed a not significant 
indication of centrality dependence, 
due to the occurrence of negative polarization, 
as well as large error bars 
\cite{BIA07,FPWW16}. 
The result of opposite directed global polarization at different 
centralities would be weird, 
if we assume that polarization comes from 
the angular momentum. Besides,  
no experimental $\Lambda$ polarization measurements, 
previous to the present ones had observed 
the opposite pointing direction of global 
$\Lambda$ polarization \cite{BHM76,Lea75,Anea84,AMS87,EJR94,AM93}. 
This might be  because of the inappropriate choice of momentum space. 
However, from the Figs. 5 and 7 in Ref. \cite{BIA07,FPWW16} 
one could still see that, the polarization signal becomes stronger 
at larger centrality, while at small centrality percentage
(below 40\%) the signal is weak and vanishing. 
Similar behavior occurs in our simulation results for 62.4 GeV, 
specifically the polarization value also vanishes when the 
centrality percentage goes below 20\%, and increases as 
the centrality increases.

The recently reported global $\Lambda$ polarization 
observation in STAR's BES I program has shown a 
positive signal for both $\Lambda$ and $\bar{\Lambda}$, 
thus it is promising to eliminate the disturbing  
opposite polarization direction that occurred in previous experiments 
\cite{BHM76,Lea75,Anea84,AMS87,EJR94,AM93}, 
and this confirm our predictions. 
Besides, the RHIC's Event Plane Detector (EPD) on upgrading for 
future BES II with higher EP resolution, will provide a better 
chance to determine the issue of centrality dependence 
of $\Lambda$ polarization \cite{JLZ2016}.
With experimental CM identification one could also verify the
momentum dependence of the polarization as shown in Fig. \ref{F1}.

%
%

\subsection{Energy Dependence and Time evolution}

The $\Lambda$ polarization increases with its
Feynman-$x_F = p_L/\sqrt{s}$, as well as transverse 
momentum $p_T$, had been observed in experiments and 
can be partly attributed to the $s \bar{s}$ pair production mechanism. 
It was also predicated that the polarization 
should also depend on the collision energy
$\sqrt{s}$, although early experiments did 
not find evident signals to confirm this \cite{AMS87,EJR94,IA06}. 
Recently with an exploration to low energy domain between
7.7 GeV to 27.0 GeV, 
the RHIC BES I program had successfully observed 
the energy dependence of $\Lambda$ 
polarization with a higher EP resolution and 
better background extraction. 

\begin{figure}[ht] 
\begin{center}
      \includegraphics[width=7.4cm]{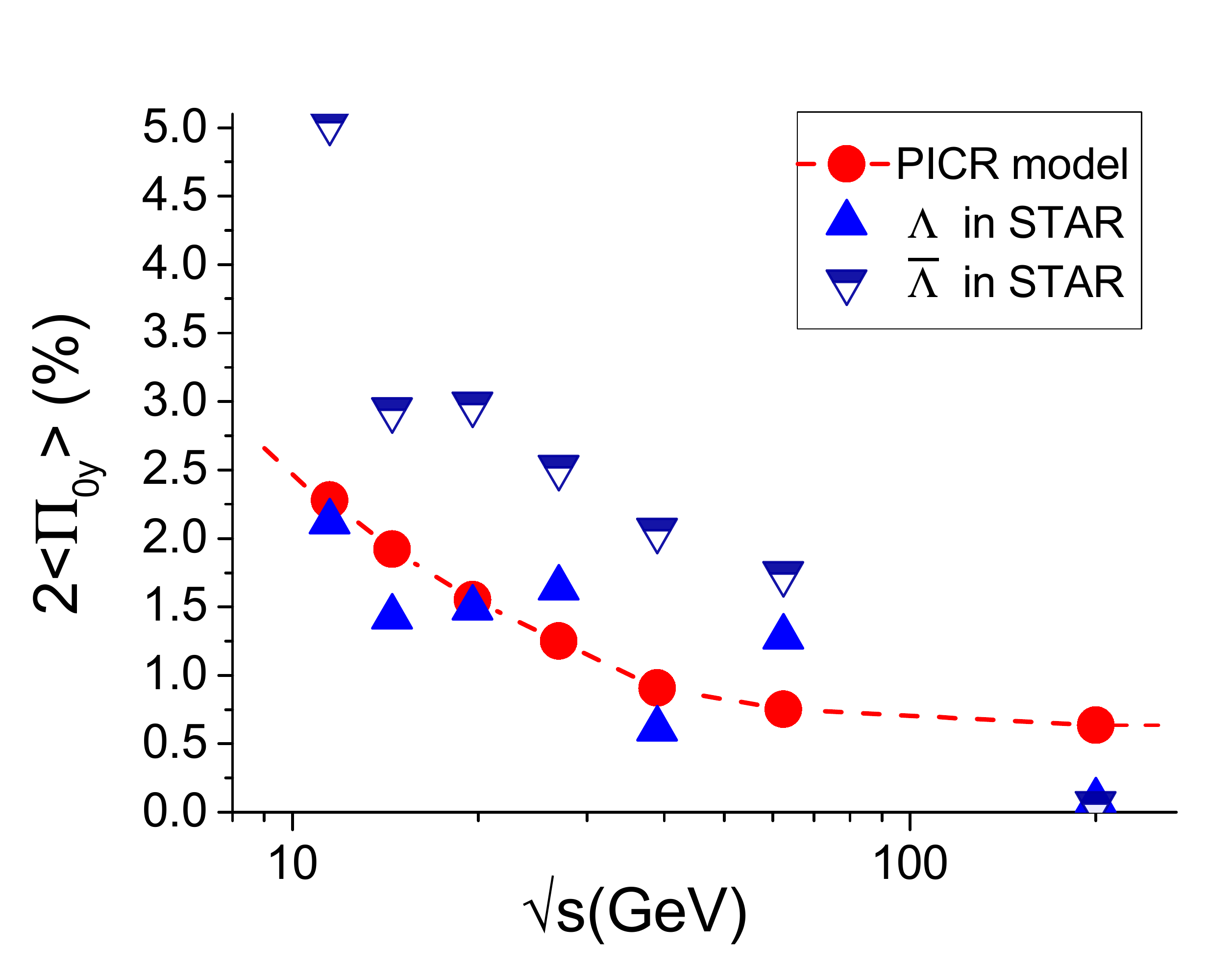}
\end{center}
\caption{
(Color online)
The global polarization, 
$2\langle\Pi_{0y}\rangle_{p}$, 
in our PICR hydro-model (red circle) and 
STAR BES experiments (green triangle),
at energies $\sqrt{s}$ of 11.5 GeV, 14.5 GeV, 19.6 GeV, 
27.0 GeV, 39.0 GeV, 62.4 GeV, and 200 GeV.
The experimental data were extracted from Ref. \cite{Lisa2016},
with solid triangle for $\Lambda$ and 
hollow triangle for $\bar{\Lambda}$,
dropping the error bars.
}
\label{F4}
\end{figure}

Using the PICR hydrodynamical model, 
we calculated the global $\Lambda$ polarization 
at the following energies: 11.5 GeV, 14.5 GeV, 19.6 GeV, 
27 GeV, 39 GeV, 62.4 GeV, and 
200 GeV, and plotted them with red round symbols in Fig. \ref{F4}. 
The impact parameter is $b_0=0.7$, i.e. the centrality is 
$c = 49\%$. For comparison the data of $\Lambda$ and 
$\bar{\Lambda}$ polarization from STAR (RHIC) were 
inserted into Fig. \ref{F4} with blue 
triangle symbols. One could see that our model 
fits fairly well the experimental data. 
Although the experimental $\bar{\Lambda}$ polarization is larger 
than the $\Lambda$ polarization, it will not change the 
averaged polarization very much, because the 
production ratio of $\bar{\Lambda}$ to $\Lambda$ is very 
small in high energy collisions \cite{TA04}. 

Fig. \ref{F4} clearly shows that $\Lambda$ polarization is 
dependent on collision energy; it drops very fast with increasing 
energy from 11.4 GeV to 62.4 GeV, and tends to saturate after 62.4 GeV. 
From thermodynamical perspective, the polarization decreases 
with energy, and this can be attributed to the higher 
temperature in higher energy collisions. The drastic 
thermal motion of particles will decrease the quark polarization rate, 
which   according to Ref. \cite{HHW11} is inversely proportional to
the collision energy. One the other hand, simulating results 
by AMPT has shown that the averaged classical vorticity 
decreases with the collision energy \cite{JLL16,DH16}, 
thus of course leads to the decline of 
global $\Lambda$ polarization.

\begin{figure}[ht] 
\begin{center}
      \includegraphics[width=7.4cm]{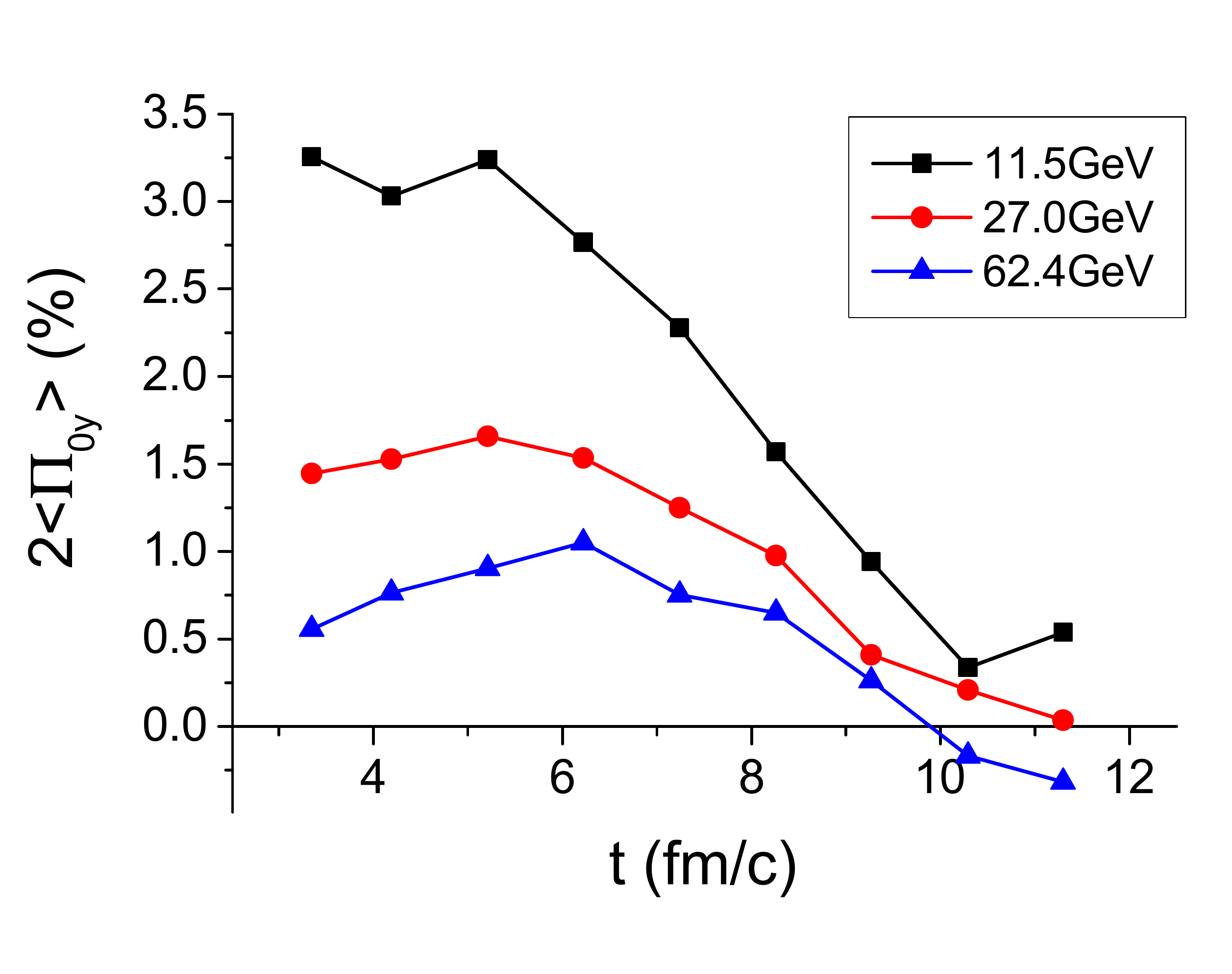}
\end{center}
\caption{
(Color online)
The time evolution of global polarization, 
$2\langle\Pi_{0y}\rangle_{p}$, 
for energy $\sqrt{s}$= 11.5 GeV, 27 GeV and 62.4 GeV.
}
\label{F5}
\end{figure}

It is also interesting to take a glance on the time 
evolution of $\Lambda$ 
polarization, shown in Fig. \ref{F5}. In this 
figure, the $\Lambda$ polarization 
increases slowly at early stage, then falls down 
very fast.
The negative polarization values that occur
at 62.4 GeV after 10 fm/c, demonstrate the 
loss of validity of the hydrodynamical model at
late stages of system expansion, due to 
the large surface to volume ratio. 
Besides, at early stages, no $\Lambda$s are 
produced, so the climbing segment of the 
curves before 4 fm/c is not observable.

%

\section{Summary and Conclusions}

With a Yang-Mills field initial state and 
a high resolution (3+1)D Particle-in-Cell 
Relativistic (PICR) hydrodynamics simulation, 
we calculate the $\Lambda$ polarization for 
different low energies and different 
impact parameters.
The polarization in high energy collisions 
originates from initial angular momentum,
or the inequality of local density between projectile and target, 
and both of them are sensitive to the impact parameter.
Thus, we plotted the global polarization as a function 
of impact parameter $b$ and a linear dependence on $b$ was observed. 
We hope that 
after upgrading the Event Plane Detector, the STAR 
will provide a higher resolution 
EP determination and centrality, to determine precisely 
the centrality dependence of 
global $\Lambda$ polarization.

Furthermore, the global $\Lambda$ polarization in our model 
decreases very fast in low energy domain, and the decline curve 
fits very well with the recent results of Beam Energy Scan (BES) 
program launched by STAR (RHIC). This is a very exciting new founding 
which indicates the significance of thermal vorticity and 
system expansion.

Finally, the time evolution of $\Lambda$ polarization shows the 
limitation of hydrodynamical model at later stage of system expansion.

\smallskip

\smallskip

\section*{Acknowledgements}

Enlightening discussions with Mike Lisa, and 
Francesco Becattini, are gratefully acknowledged. 
One of the authors, Y.L. Xie, is supported by the China 
Scholarship Council (China).


\end{document}